\documentclass{aastex}
\usepackage{emulateapj5,epsfig}

\newenvironment{inlinefigure}{ 
\def\@captype{figure} 
\noindent\begin{minipage}{0.999\linewidth}\begin{center}} 
{\end{center}\end{minipage}\smallskip} 
\newcommand{\beq}{\begin{equation}}
\newcommand{\eeq}{\end{equation}}

\usepackage{graphicx}
\usepackage{amssymb}
\usepackage{epstopdf}
\font\tenbg=cmmib10 at 10pt

\def \rvecphi{{\hbox{\tenbg\char'036}}}

\def \trho{{\tilde{\rho}}}
\def\lesssim{\mathrel{\hbox{\rlap{\hbox{\lower4pt\hbox{$\sim$}}}\hbox{$<$}}}}
\def\gtrsim{\mathrel{\hbox{\rlap{\hbox{\lower4pt\hbox{$\sim$}}}\hbox{$>$}}}}

\begin{document}
\title{Vertical Structure of Stationary Accretion
Disks with a Large-Scale Magnetic Field}

\author{G.S. Bisnovatyi-Kogan\altaffilmark{1}
\&  R.V.E. Lovelace\altaffilmark{2}}

\altaffiltext{1}{Space Research Institute,
Russian Academy of Sciences, Moscow, Russia;
gkogan@mx.iki.rssi.ru}

\altaffiltext{2}{Department of Astronomy,
Cornell University, Ithaca, NY 14853-6801;
RVL1@cornell.edu}

\begin{abstract}

    In earlier works we
pointed out that   the disk's surface
layers  are non-turbulent and
thus highly conducting (or non-diffusive)
because the hydrodynamic and/or  magnetorotational (MRI)
instabilities are suppressed
high in the disk  where    the magnetic
{and radiation pressures}
are  larger than the plasma thermal pressure.
     Here, we calculate the vertical profiles
of the {\it stationary} accretion flows (with radial
and azimuthal components), and
the profiles of the  large-scale, magnetic field taking into
account the turbulent viscosity and diffusivity
 and the fact that the turbulence
vanishes at the surface of the disk.
    Also, here we require that the
radial accretion speed be zero at the disk's
surface and we assume that the ratio of the turbulent
viscosity to the turbulent magnetic diffusivity is
of order unity.   
    Thus at the disk's surface there are three boundary conditions.
      As a result, for a fixed dimensionless viscosity
$\alpha$-value, we find that there is a definite relation between
 the ratio ${\cal R}$ of the accretion power going into magnetic disk winds
 to the viscous power dissipation 
and the midplane plasma-$\beta$, which is the ratio of
the plasma to magnetic pressure in the disk.

   For a specific disk model with ${\cal R}$ of order unity
we find that  the critical value required for a stationary solution is
$\beta_c \approx 2.4r/(\alpha h)$, where $h$ the disk's half thickness.
For weaker magnetic fields,
$\beta > \beta_c$, we argue that the poloidal field will
advect outward while for $\beta< \beta_c$ it will advect
inward.   Alternatively, if the disk wind is negligible
(${\cal R} \ll 1$), there are stationary solutions with $\beta \gg
\beta_c$.

\end{abstract}


\section{Introduction}

         Analysis of the diffusion and advection of a
large-scale magnetic field in a  accretion disk with a turbulent
viscosity and magnetic diffusivity arising from the 
magnetorotational instability (MRI)
shows  that a {\it weak} large-scale field 
diffuses outward rapidly (van Ballegooijen 1989;
Lubow, Papaloizou, \& Pringle 1994).  
      We mention but do not consider here the opposite limit 
where the magnetic field is sufficiently strong
that it suppresses  the MRI instability so that
the disk is non-turbulent but accretion occurs due
to angular momentum outflow to a magnetic
disk wind or jet (Lovelace, Romanova, \& Newman 1994).
       Earlier,  Bisnovatyi-Kogan and Lovelace (2007)
pointed out that  the disk's surface
layers are highly conducting 
because  the MRI instability  is suppressed
in this region where  the
magnetic and radiative energy-densities are larger than
the thermal gas  energy-density.
   Rothstein and Lovelace (2008)  analyzed this problem
in further detail and discussed the connections
with global and shearing box magnetohydrodynamic
(MHD) simulations of the MRI.
Lovelace, Rothstein, \& Bisnovatyi-Kogan
(2009; hereafter LRBK) developed an analytic model for the vertical
($z$)      profiles of the stationary accretion flows (with radial
and azimuthal components), and
the profiles of the  large-scale,
magnetic field taking into
account the turbulent viscosity and diffusivity
due to the MRI and the fact that the turbulence
vanishes at the surface of the disk.

     Here, we require that the
radial accretion speed be zero at the disk's
surface, and we assume that the ratio of the turbulent
viscosity to the turbulent magnetic diffusivity is
of order unity as suggested by MHD shearing-box
simulations (Guan \& Gammie 2009).
     For a fixed dimensionless viscosity
$\alpha$-value, we find that there is a definite relation between
 the ratio ${\cal R}$ of the accretion power going into magnetic disk winds
 to the viscous power dissipation 
and the midplane plasma-$\beta$, which is the ratio of
the plasma to magnetic pressure in the disk.

    Section 2 discusses the model
for the flow and ordered magnetic field in
a viscous diffusive disk.   Section 3 discusses the
solutions for a specific disk model.
Section 4 gives the conclusions.

\section{Theory}

       Following LRBK we consider the non-ideal magnetohydrodynamics
of a thin axisymmetric, viscous, resistive disk threaded
by a large-scale dipole-symmetry magnetic field ${\bf B}$.
  We use a cylindrical $(r,\phi,z)$ inertial coordinate system
in which the time-averaged
magnetic field is ${\bf B}=B_r\hat{\bf r}+B_\phi\hat{\rvecphi~}+
B_z\hat{\bf z}$, and the time-averaged flow velocity is
${\bf v}=v_r\hat{\bf r}+v_\phi\hat{\rvecphi~}+
v_z\hat{\bf z}$.
    The main equations are
\begin{eqnarray}
\rho {d{\bf v}\over dt}&=&-{\bf \nabla} p +\rho{\bf g}
+ {1\over c}{\bf J \times B} + {\bf F^\nu}~,
\\
{\partial {\bf B} \over \partial t}
&=& {\nabla \times}({\bf v \times B}) -
{\bf \nabla}\times(\eta {\bf \nabla }\times {\bf B})~.
\end{eqnarray}
These equations are supplemented  by the continuity equation,
$\nabla\cdot(\rho {\bf v})=0$, by
${\bf \nabla \times B} =4\pi{\bf J}/c$, and by
${\bf\nabla} \cdot {\bf B}=0$.
   Here, $\eta$ is the magnetic
diffusivity, ${\bf F}^\nu=-{\bf \nabla}\cdot T^\nu$
is the viscous force with
$T_{jk}^\nu= -\rho \nu (\partial v_j/\partial x_k
+\partial v_k/\partial x_j-(2/3)\delta_{jk}
{\bf \nabla} \cdot{\bf v} )$ (in Cartesian
coordinates), and $\nu$ is the
kinematic viscosity.
     For simplicity,  in place of an energy equation
we consider the adiabatic dependence $p \propto \rho^\gamma$,
with $\gamma$ the adiabatic index.

   We assume that both the viscosity and
the diffusivity are
due to  magneto-rotational (MRI) turbulence in
the disk  so that
\begin{equation}
\nu ={\cal P} \eta =\alpha ~{c_{s0}^2 \over \Omega_K}~ g(z)~,
\end{equation}
where ${\cal P}$ is the magnetic Prandtl number of
the turbulence assumed
a constant of order unity (Bisnovatyi-Kogan \& Ruzmaikin 1976),
$\alpha \leq 1$ is the  dimensionless Shakura-Sunyaev
(1973) parameter, $c_{s0}$ is the midplane isothermal sound
speed, $\Omega_K \equiv
(GM/r^3)^{1/2}$ is
the Keplerian angular velocity
of the disk, and $M$ is the
mass of the central object.
     The function  $g(z)$
accounts for the absence of turbulence in
the surface layer of the disk (Bisnovatyi-Kogan \& Lovelace
2007;  Rothstein \& Lovelace 2008).
    In the  body of the disk $g = 1$, whereas
at the surface of the disk, at say $z_S$, $g$ tends over
a short distance to
a very small value
$\sim 10^{-8}$, effectively zero, which is the ratio of the Spitzer
diffusivity of the disk's
surface layer to the turbulent diffusivity of
the body of the disk.
 At the disk's surface the density is much
 smaller than its midplane value.

     We consider stationary solutions of equations (1) and (2)
for a weak large-scale magnetic field.
These can be greatly simplified
for thin disks where the disk half-thickness, of
the order of  $h \equiv c_{s0}/\Omega_K$,
is much less than $r$.
Thus we have the small parameter
\begin{equation}
\varepsilon={h \over r} = {c_{s0}\over v_K} \ll 1~.
\end{equation}
It is useful in the following to use the
dimensionless height $\zeta \equiv z/h$.
      The midplane plasma beta is taken to be
\begin{equation}
\beta \equiv {4\pi \rho_0c_{s0}^2 \over B_0^2}~,
\end{equation}
where $\beta =c_{s0}^2/v_{A0}^2$,
$v_{A0}=B_0/(4\pi\rho_0)^{1/2}$ is the midplane
Alfv\'en velocity.  Note that the conventional
definition of beta is $2\beta$.
     The rough condition for the MRI instability
and the associated turbulence in the disk is ${\beta} \gtrsim 1$
(Balbus \& Hawley 1998) and this is assumed here.

  The three magnetic field components are  assumed
to be of comparable magnitude on the disk's surface,
but $B_r = 0 = B_\phi$ on the midplane.
  On the other hand the axial magnetic field changes
by only a small almount going from the midplane
to the surface, $\Delta B_z \sim \varepsilon  B_r \ll B_z$
(from ${\bf \nabla} \cdot{\bf B} =0$) so that $B_z\approx$ const inside
the disk.
   As a consequence, the
$\partial B_j/\partial r$ terms in the
magnetic force  in equation (1) can all be dropped
in favor of  the $\partial B_j/\partial z$ terms
(with $j=r,~\phi$).
   It is important to keep in mind that $B_j$ is the
large scale field;  the approximation does not
apply to the small-scale field which gives
the viscosity and diffusivity.
   The three velocity components are
assumed to satisfy $v_z^2 \ll c_{s0}^2$
and  $v_r^2 \ll v_\phi^2$.
Consequently,  $v_\phi(r,z)$ is  close in
value to the Keplerian value $v_K(r)\equiv (GM/r)^{1/2}$,
except in the outer disk layers where the radial magnetic force
may be comparable with the centrifugal and gravitational forces.
   We normalize the field components by $B_0 = B_z(r,z=0)$,
with  $b_r=B_r/B_0$, $b_\phi = B_\phi/B_0$,
and $b_z = B_z/B_0\approx 1$.
   Also, we define $u_\phi \equiv v_\phi(r,z)/v_K(r)$ and
the accretion speed $u_r \equiv -~v_r/(\alpha c_{s0})$.
  For the assumed dipole field symmetry, $b_r$ and $b_\phi$
are odd functions of $\zeta$ whereas $u_r$ and $u_\phi$
are even functions.

   Integration over the vertical extent of the disk gives
the average accretion speed
\begin{equation}
\overline{u}_r = u_0 - {2 b_{\phi S+}\over
\alpha \beta \tilde{\Sigma}}~,
\end{equation}
(LRBK) which is the sum of a viscous
contribution, $u_0\equiv 3\varepsilon k_\nu$ (with $k_\nu$
a numerical constant of order unity), and a magnetic
contribution ($\propto b_{\phi S+}$) due to the loss of angular
momentum from the surface of the disk where
necessarily $b_{\phi S+} \leq 0$ (Lovelace,
Romanova, \& Newman 1994).

     We assume $p \propto \rho^\gamma$
so that the vertical hydrostatic equilibrium
gives
\begin{equation}
\tilde{\rho}={\rho \over \rho_0} =\left(
1-{(\gamma-1)\zeta^2 \over 2\gamma}\right)^{1/(\gamma-1)}~,
\end{equation}
for $\beta \gg 1$.
    The density goes to zero at
$\zeta_m = [2\gamma/(\gamma-1)]^{1/2}$.
   However, before this distance is reached the
MRI turbulence is suppressed, and $g(\zeta)$ in
equation (3) is effectively zero.

     The different components of equations (1)
and (2) can be combined (LRBK) to give the
following equation for the radial accretion speed,
\begin{eqnarray}
{\alpha^4\beta^2}
{\partial^2 \over \partial \zeta^2}
\left(g{\partial \over \partial \zeta}
\left(\trho g{\partial \over \partial \zeta}
\left({1\over \trho}{\partial \over \partial \zeta}
\left(\trho g {\partial u_r \over \partial \zeta}\right)\right)
\right)\right)
\nonumber \\
-~\alpha^2\beta {\cal P}
{\partial^2 \over \partial \zeta^2}
\left(g {\partial \over \partial \zeta}
\left(\trho g{\partial \over \partial \zeta}
\left({u_r \over \trho g}\right)\right)\right)
\nonumber\\
-~\alpha^2\beta {\cal P}
{\partial^2 \over \partial \zeta^2}\left({1\over \trho}
{\partial \over \partial \zeta}
\left(\trho g {\partial u_r \over \partial \zeta}\right)\right)
\nonumber \\
+~{\alpha^2\beta^2 }
{\partial^2 \over \partial \zeta^2}\bigg(\trho g \big(u_r- gu_0
\big)\bigg)
+{\cal P}^2 {\partial^2 \over \partial \zeta^2}\bigg({u_r \over \trho g}
\bigg)
\nonumber \\
+~3\beta{\cal P }^2{u_r \over g}=0~.
\end{eqnarray}
  The equation can be integrated from $\zeta =0$
out to the surface of the disk $\zeta_S$ where
boundary conditions apply.

     For specificity we take
$g(\zeta)=(1-{\zeta^2/ \zeta_S^2})^\delta ,$
where
 $\zeta_S  <\zeta_m$ and $\delta$ is a constant.
    That is, we neglect the ratio of the Spitzer
diffusivity on the surface of the disk  to
its value in the central part of the disk.
  An  estimate of $\zeta_S$ can be made
by noting that $\beta(\zeta)=4\pi p(\zeta)/B_0^2 =
\beta (\tilde{\rho})^\gamma \approx 1$
at $\zeta_S$.
    This gives $\zeta_S^2/\zeta_m^2=
1-\beta^{-(\gamma-1)/\gamma}$ and
$\rho_S/\rho_0 = \beta^{-1/\gamma}$.

\subsection{Boundary Conditions}

   We consider only the solutions
which  have  net mass accretion,
$
\dot{M}=4\pi r h \rho_0 \alpha c_{s0}\tilde{\Sigma}~
\overline{u}_r > 0$,
and  have $b_{\phi }\leq 0$ on the disk's surface.
   This condition on $b_{\phi S+}$ corresponds to an
efflux of angular momentum and energy
(or their absence) from
the disk to its corona rather than the reverse.
     The condition $b_{\phi S+} \leq 0$
 is the same as $\overline{u}_r \geq  u_0$,
 where $u_0$ is the minimum (viscous) accretion speed.
   Note that ${\cal R}=\overline{u}_r/u_0 -1$ is the ratio of the
accretion power which goes into the disk winds to the
viscous power dissipation.
    Clearly, the condition on $b_{\phi S+}$ implies that $\dot{M}>0$
so that there is  only one condition.
    In general there is a continuum of values of $b_{\phi S} \leq 0$
for the considered solutions inside the disk.
   The value of $b_{\phi S }$ can be determined by matching
the calculated fields $b_{rS}$ and $b_{\phi S}$ onto
an external field and flow as discussed in LRBK and here
in \S 3.

    LRBK showed that there is no
jump in $b_r$ across the conducting surface layer.
     This  implies that
$
{\partial u_r/ \partial \zeta}|_{\zeta_S} =0,$
which represents a first boundary condition of the solution
of equation (8).
The equations inter-reltaing $b_r$, $b_\phi$, $u_r$, and $\delta u_\phi$
are summarized in the Appendix.
   A second boundary condition,
$
u_r |_{\zeta_S} =0,$
follows from equation 12
of LRBK evaluated just outside of the conducting layer.
     A third boundary condition,
$
{\partial b_\phi / \partial \zeta} {|}_{\zeta_{S-}} =0,$
is derived in LRBK.

\begin{inlinefigure}
\centerline{\epsfig{file=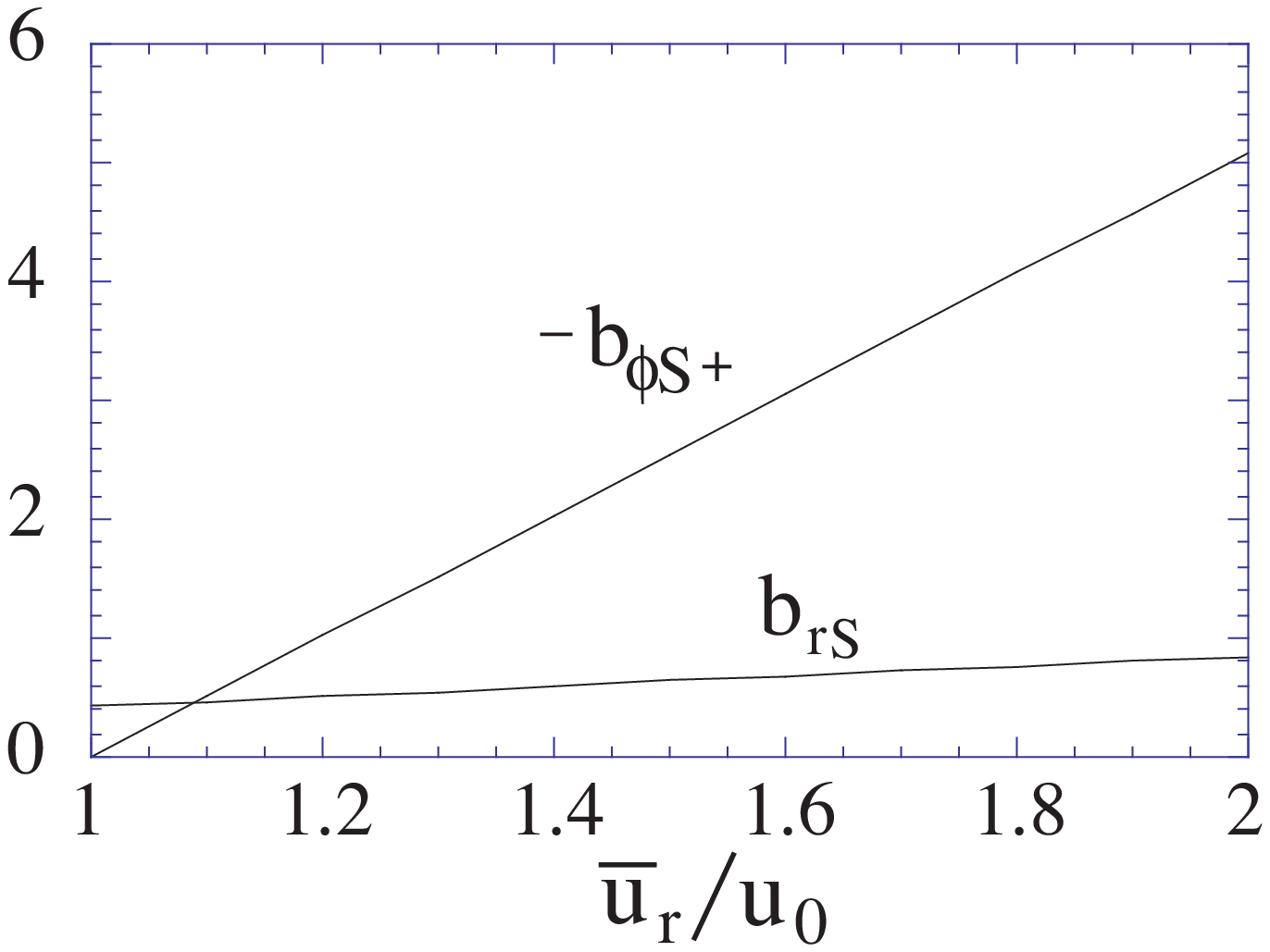,
height=2.2in,width=3in}}
\epsscale{0.5}
\figcaption{Radial and toroidal field components
normalized to $B_z(r,0)$ at the disk's surface
as a function of the average accretion speed
$\overline{u}_r$ (normalized by the
viscous accretion speed $u_0$).
   For this plot
$\beta=240$ and
${\cal P}=1$. }
\end{inlinefigure}

 \section{Specific Solutions}

    Here, to simplify the analysis we
consider the limit where
$\gamma \rightarrow  \infty$ in equation
(7) and $\delta \rightarrow 0$ in
equation (18) of LRBK.
   Then, $\zeta_S \rightarrow \zeta_m$ and
both $\tilde{\rho}$ and $g$
are unit step functions going to zero at $\zeta_m =\sqrt{2}$.
Also,  $\overline{u}_r = \langle u_r \rangle$ and $\tilde{\Sigma}=
\sqrt{2}$.
   Thus the above physical condition
$\overline{u}_r \geq 3 \varepsilon k_\nu=u_0$ implies
that $b_{rS} \geq u_0 \zeta_S{\cal P}$
from equation (13) of LRBK.   We assume $k_p \equiv - \partial \ln p/\partial \ln r$ and $k_\nu =1$.

   The solutions to equation (8) are
$u_r \propto \exp(ik_j \zeta)$ (with $j=1,~2,~3$), where
\begin{equation}
 \alpha^4\beta^2(k_j^2)^3+2{\cal P}\alpha^2\beta(k_j^2)^2+
 (\alpha^2 \beta^2+{\cal P}^2)k_j^2-3\beta{\cal P}^2  =0~,
\end{equation}
is a cubic in $k_j^2$.
  The discriminant of the cubic is negative so that there
is one real root, $k_1^2$, and a complex conjugate
pair of roots,
$k_{2,3}^2$.
    Because $u_r$ is an even function of $\zeta$ we
can write
\begin{eqnarray}
u_r& = &
a_1 \cos(k_0\zeta)+a_2 \cos(k_r\zeta) \cosh(k_i\zeta)
\nonumber\\
&+&a_3 \sin(k_r\zeta)\sinh(k_i\zeta)~,
\end{eqnarray}
where $k_0 =\sqrt{k_1^2}$,
$k_r ={\rm Re}(\sqrt{k_2^2})$, and
$k_i ={\rm Im}(\sqrt{k_2^2})$.

      We consider a thin disk,
$\varepsilon =h/r=0.05$,  and  a
viscosity parameter $\alpha=0.1$.
     Figure 1 shows the dependences of
the surface field components on the
average accretions speed for $\beta=240$
and  ${\cal P}=1$.
The $b_{\phi S}$ dependence is given by
equation (6) and is independent of ${\cal P}$
while the $b_{rS}$ is given by $b_{rS} =
{\cal P} \zeta_S \overline{u}_r$ from
LRBK (equation 13).

\begin{inlinefigure}
\centerline{\epsfig{file=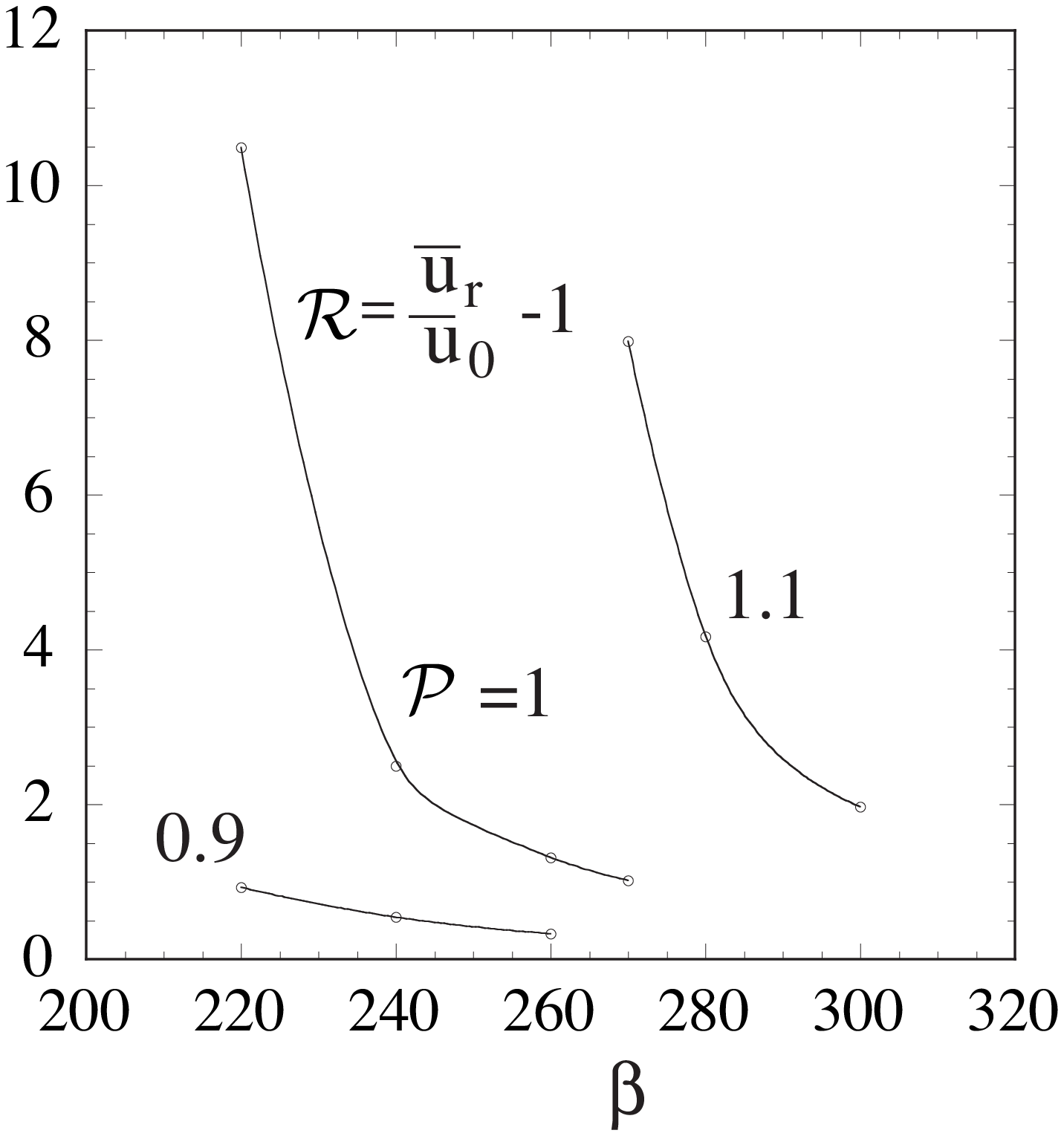,
height=2.8in,width=3.1in}}
\epsscale{0.5}
\figcaption{Summary of solutions to equation (8)
which satisfy the boundary conditions of \S 2.1.
$\overline{u}_r$ is the average accretion speed and
$u_0$ is the accretion speed in the absence of an
ordered magnetic field.   The curves are smooth
lines connecting the marked points. The physically relevant solutions 
have  ${\cal R} \lesssim {\cal O}[2\sqrt{2}\pi/(\alpha \beta u_0)]$
which is about $2.5$ for $\alpha=0.1,~\beta=240,$ and $u_0=0.15$.}
\end{inlinefigure}

   There are three boundary conditions at the surface of the
disk.  Additionally, there is the  ratio ${\cal R}$ of the accretion power going into magnetic disk winds to the viscous power dissipation.
      The solution for $u_r(\zeta)$
has three unknown coefficients, $a_1$, $a_2$.
\& $a_3$ which are dependent on $\alpha$, ${\cal P}$,
and $\beta$.
    If  $\alpha$ and ${\cal P}$ are fixed, then there is a definite relation between ${\cal R}$ and $\beta$.
     Figure 2 shows this relation determined numerically.

\begin{inlinefigure}
\centerline{\epsfig{file=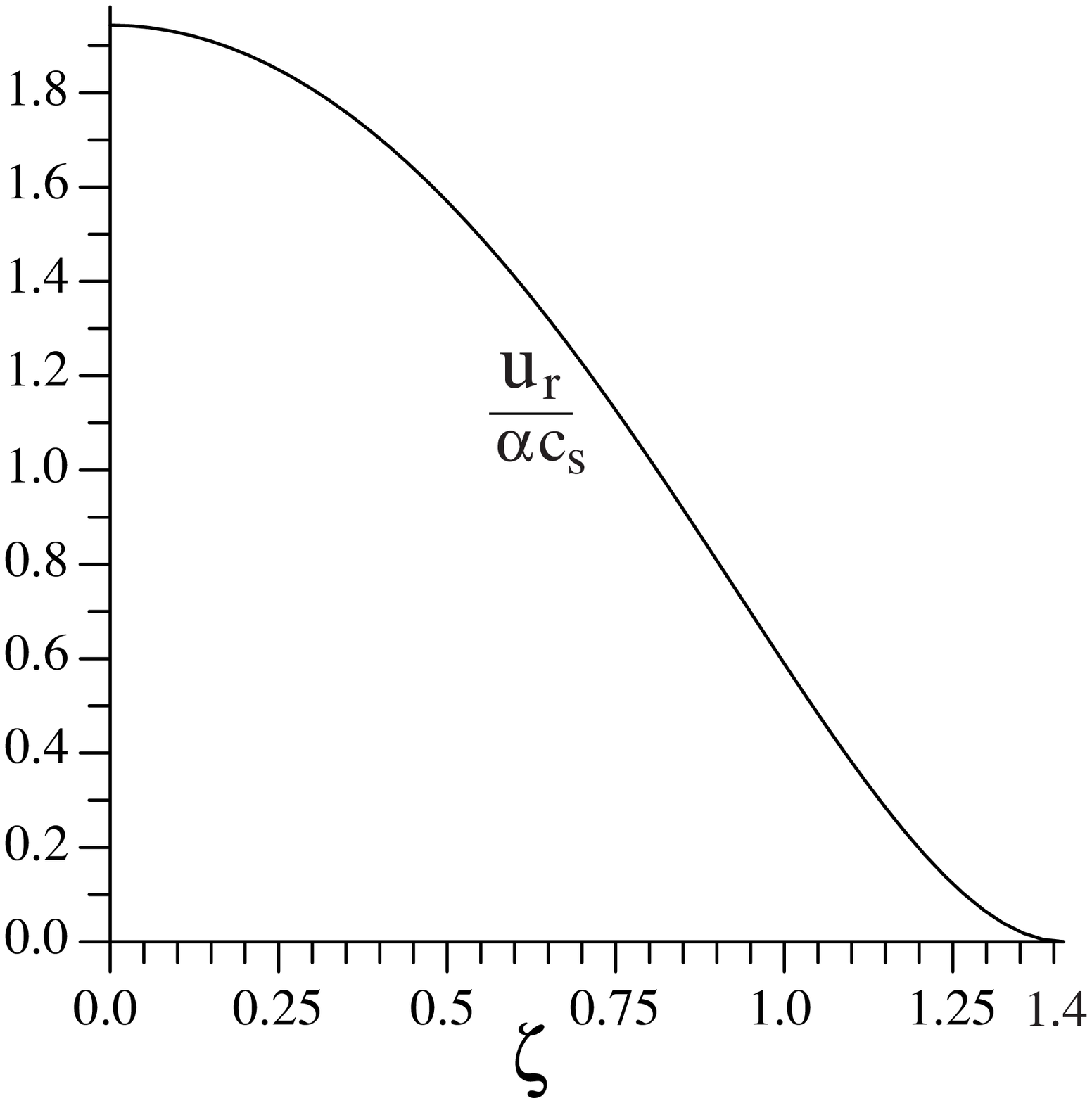,
height=2.5in,width=2.9in}}
\epsscale{0.5}
\figcaption{Radial inflow speed $u_r = -v_r$ (normalized
to $\alpha c_{s0}$) as a function
of $\zeta=z/h$ for $\beta=240$ and ${\cal P}=1$. 
The surface of the disk is at $\zeta=\sqrt{2}$.}
\end{inlinefigure}

    Figures 3 shows the vertical profile of the accretion flow
 for $\beta=240$.  The profiles of $\delta u_\phi=
 (v_\phi-v_K)/v_K$,  $b_r$, $b_\phi$ follow from 
 the equations in the Appendix.

     The value of $b_{\phi S+}\leq 0$ or $\overline{u}_r/u_0$ is not
fixed by the solution for the field and flow inside the disk.
     Its value can be determined by matching the calculated
surface fields $b_{r S}$  and $b_{\phi S+}$ onto an external
magnetic wind or jet solution.   Stability of the wind or jet
solution to current driven kinking is predicted to
limit the ratio of the toroidal to axial magnetic field
components at the disk's surface $|b_{\phi S +}|$ to values
 $\lesssim {\cal O}(2\pi)$ (Hsu \& Bellan 2002; Nakamura, Li,
 \& Li 2007).
      Recall that ${\cal R}=\overline{u}_r/u_0 -1= 2|b_{\phi S +}|/
(\alpha \beta \sqrt{2}  u_0)$ is
the ratio of the accretion power going into the disk wind to
the viscous dissipation in the disk.
     For the mentioned upper limit on $|b_{\phi S+}|$,
${\cal R}\lesssim {\cal O}[2\sqrt{2}\pi/(\alpha \beta u_0)] \approx 2.5$.
      This implies a critical value of $\beta$
for a stationary disk solution (for ${\cal P}=1$),
$\beta_c \approx 1.2(\alpha \varepsilon)^{-1}$.
   For $\alpha=0.1$ and $\varepsilon=0.05,$ $\beta_c \approx 240$.

     For $\beta \lessgtr \beta_c$,  the disk and large-scale magnetic field is not be in a stationary state.
      Equation (6) for the average accretion speed still
applies and can be written as $\overline{u}_r = u_0 + u_B$
where $u_B$  is the magnetic contribution to the accretion due
to the outflow of angular momentum from the disk's surfaces.
   In general, $u_B$ is an increasing function of the magnetic
field strength if ${\cal R} \sim 1$ (Lovelace et al. 1994).
    For this reason for $\beta < \beta_c$ the ordered
poloidal magnetic field threading the disk will be advected inward in the disk while for $\beta > \beta_c$  the field will advect outward.
    Thus, $\beta_c$ acts as a threshold value for the  buildup
of a large scale field in the inner regions of an accretion disk.

       Another possibility is that the disk winds are negligible so
that  ${\cal R} \ll 1$.   Figure 2 suggests that in this limit we
have a stable stationary solutions with $\beta$ values much
larger than the mentioned critical value.

\section{Conclusions}

   For a specific  disk model with  turbulent Prandtl
number ${\cal P}=1$ and ${\cal R} \sim 1$,we  show that there is  a critical value
${\beta}_c\approx
1.2r/(\alpha h)$ for a {\it stationary} disk threaded by a  large-scale magnetic field.
  For $\alpha=0.1$ and $\varepsilon=0.05$, ${\beta}_c\approx 240$.
    For ${\beta} < {\beta}_c$,  we argue that
that the large-scale field will be advected inward and lead to
the build up of the ordered field in the central region of the disk.
   For ${\beta}>{\beta}_c$, the ordered field will
advect outward.    Alternatively, if the disk wind is negligible
so that ${\cal R} \ll 1$, stationary disk equilibria with a large
scale field are possible for $\beta$ much larger than the mentioned
$\beta_c$.

       The 3D MHD simulations of Beckwith,
Hawley, \& Krolik (2009)  show the inward advection
of a significant fraction of the initial unipolar
vertical magnetic flux threading
a MRI unstable plasma torus around a black hole.
  The initial average  ${\beta}$ in the torus is $100$ which
is significantly less than the critical value estimated here.
   A broader range of simulations for much larger initial values
of $\beta$ would be needed to determine if the initial
flux diffuses outward.  Axisymmetric MHD simulations of
Romanova et al. (2011) show episodes of enhanced accretion
on the disk surfaces for the case of an MRI unstable
disk around  a rotating  magnetized stars.   However, the presence of  the stellar magnetic field complicates the analysis of flux transport in the disk.

We have shown that the presence of a large scale 
poloidal magnetic field in an accretion disk 
around a black hole not only
plays a decisive role in the jet formation,  particle acceleration, and observed hard energy radiation (Bosch-Ramon 2011), 
but also creates the possibility for a new type of
non-stationary behavior in these objects, which could be related to appearance of different spectral and luminosity
states and transitions between them as observed in microquasars (Malzac and Belmont 2009).

\section*{Acknowledgments}

The work of G.S.B.-K. was partially supported by RFBR grants
08-02-00491 and 08-02-90106, RAN Program ``Formation
and evolution of stars and galaxies.''
R.V.E.L  was supported in part by NASA grants
NNX10AF63G and NNX11AF33G and by NSF grant AST-1008636.

\appendix

\section{Subsidiary Equations}

   Here, we summarize the relations between the field and velocity
components.   From the radial force balance,
\begin{equation}
{\partial b_r  \over \partial \zeta}
= {\beta \tilde{\rho}\over \varepsilon} ~~
\big(1 - k_p~\varepsilon^2 - u_\phi^2 \big) +\alpha^2\beta
{\partial \over \partial \zeta}
\left(\tilde{\rho} g{\partial u_r \over \partial \zeta} \right)~,
\end{equation}
where $k_p \equiv -d \ln p/d\ln r$ is assumed positive and of
order unity.
     The $\phi-$component of equation (2)
gives
\begin{equation}
{\partial b_\phi \over \partial \zeta}
={ \alpha \beta \tilde{\rho}\over 2} ~
\big( 3 \varepsilon k_\nu g - u_r \big)
-{\alpha \beta \over \varepsilon}
{\partial \over \partial \zeta}\left(\tilde{\rho}g
{\partial u_\phi \over \partial \zeta}\right)~,
\end{equation}
where $k_\nu \equiv d \ln(\rho c_{s0}^2 r^2/h)/ d \ln(r)>0$
is also assumed to be of order unity.

The toroidal component of Ohm's law  gives
\begin{equation}
{\partial b_r \over \partial \zeta} ={{\cal P}\over g} {u_r }~.
\end{equation}
The other components of Ohm's law give
\begin{equation}
{\partial u_\phi \over \partial \zeta}
={3\varepsilon \over 2} b_r -
{\alpha \varepsilon \over {\cal P}}{\partial \over \partial \zeta}
\left( g
{\partial b_\phi \over \partial \zeta}\right)~.
\end{equation}

   Combining equations (A1) and (A3) gives
\begin{equation}
u_r ={\beta \tilde{\rho} g \over \varepsilon{\cal P}}
\big(1-k_p\varepsilon^2 - u_\phi^2\big)+{\alpha^2\beta g\over {\cal P}}
{\partial \over \partial \zeta}
\left(\trho g{\partial u_r \over \partial \zeta}\right)~.
\end{equation}
For thin disks, $\varepsilon \ll 1$, and $\beta > 1$,
we have $u_\phi = 1 +\delta u_\phi$ with
$(\delta u_\phi)^2 \ll 1$.
      Consequently,
\begin{equation}
\delta u_\phi = -{k_p\varepsilon^2\over 2}
-{\varepsilon {\cal P}u_r \over 2\beta \tilde{\rho}g}
+{\alpha^2 \varepsilon \over 2 \tilde{\rho}}
{\partial \over \partial \zeta}
\left(\tilde{\rho}g {\partial u_r \over \partial \zeta}\right)~,
\end{equation}
 to a good approximation.

\end{document}